\documentclass[conference,11pt]{IEEEtran}
\usepackage{cite}
\usepackage{soul}
\usepackage{amsmath,amssymb,amsfonts}
\usepackage{algorithmic}
\usepackage{graphicx} 
\usepackage{textcomp}
\usepackage{xcolor}
\usepackage{hyperref}
\usepackage{subfigure}
\usepackage[normalem]{ulem}
\usepackage{comment}

\title{Neural Field-of-View for Binaural Signal Matching with Wearable Microphone Arrays}
\author{\IEEEauthorblockN{Matan Yifrach\IEEEauthorrefmark{1}, Boaz Rafaely,~\textit{Fellow,~IEEE}\IEEEauthorrefmark{1}\IEEEauthorrefmark{2}} \IEEEauthorblockA{\IEEEauthorrefmark{1}School of Electrical and Computer Engineering, Ben-Gurion University of the Negev, Beer-Sheva 84105, Israel} \IEEEauthorblockA{\IEEEauthorrefmark{2}Corresponding author, email: br@bgu.ac.il}}

\begin{document}
\maketitle

\begin{abstract}
The growing use of spatial audio in applications such as augmented and virtual reality has driven the development of binaural reproduction methods for wearable arrays with a limited number of microphones. Binaural signal matching (BSM) is one such method, producing high-quality binaural signals under a diffuse-field assumption, but degrading at high direct-to-reverberant ratios (DRR) where the direct sound dominates. Previous extensions incorporate Field-of-View (FoV) weighting, either with fixed apertures or based on explicit source localization, but these approaches are limited by coarse spatial coverage or reliance on localization estimation accuracy. This paper introduces FoV-BSM-Net, a signal-dependent FoV-BSM formulation that avoids explicit source estimation by learning the FoV parameters end-to-end from the microphone signals using a Convolutional Recurrent Neural Network. The method is evaluated in simulated rooms across varying reverberation conditions, and compared against BSM and a fixed FoV-BSM baseline. Results show that FoV-BSM-Net consistently improves over BSM, with gains that grow with DRR in both binaural NMSE and interaural cue errors, and are further supported by perceptual evaluation showing a substantial advantage over both baselines across low and high DRR conditions.
\end{abstract}


\section{Introduction}
The increasing interest in immersive technologies, such as virtual reality (VR), augmented reality (AR), and high-fidelity teleconferencing, has increased the importance of binaural reproduction \cite{Richard:AudioSignalProcessing:2023,Rafaely:SpatialAudioSignal:2022b,Ben-Hur:SpectralEqualizationBinaural:2017} to recreate natural and immersive sound over headphones.
%
A common approach is to use High-Order Ambisonics (HOA), which encodes the sound field into spherical harmonic (SH) components to be convolved with head-related transfer functions (HRTFs) \cite{Moreau:3DSoundField:2006,Rafaely:InterauralCrossCorrelation:2010,Poletti:ThreedimensionalSurroundSound:2005}. While providing a high-quality binaural signal, the capture of HOA typically requires spherical arrays with a high number of microphones, which may not be available in practice. Although previous studies explored the capture of HOA with non-spherical arrays with only a few microphones~\cite{Ahrens:SphericalHarmonicDecomposition:2021,Ahrens:SphericalHarmonicDecomposition:2022,McCormack:ParametricAmbisonicEncoding:2022,Gayer:AmbisonicsEncodingArbitrary:2024a}, this remains a challenging open problem.

As an alternative to encoding HOA, previous work has studied the direct encoding of binaural signals. In Beamforming-Based Binaural Reproduction (BFBR), microphone signals are filtered with a set of beamformers steered toward multiple directions and convolved with the corresponding HRTFs to produce the binaural output \cite{ODonovan:SphericalMicrophoneArray:2008,Song:UsingBeamformingBinaural:2008,Calamia:ConformalHelmetmountedMicrophone:2017,Ifergan:SelectionNumberBeamformers:2022}.
Further studies have sought to improve binaural reproduction through improved HRTF encoding \cite{Rasumow:SmoothingHeadrelatedTransfer:2012,Schorkhuber:BinauralRenderingAmbisonic:2018,Rasumow:PerceptualEvaluationIndividualized:2017,Ben-Hur:BinauralReproductionBased:2021}, 
or by explicitly minimizing the binaural error using the binaural signal matching (BSM) technique \cite{Setsompop:MagnitudeLeastSquares:2008, Madmoni:DesignAnalysisBinaural:2024}. 

While the BSM approach, which was further improved by applying magnitude least-squares (MagLS) optimization\cite{Madmoni:DesignAnalysisBinaural:2024,Setsompop:MagnitudeLeastSquares:2008,Deppisch:EndtoEndMagnitudeLeast:2021a}, and perceptually-motivated loss functions \cite{Berebi:FeasibilityIMagLSBSMILD:2024,Berebi:BSMiMagLSILDInformed:2025a,Berebi:NeuralNetworkSolvers:2025}, represents a major advance, it is considered a signal-independent approach, assuming the sound field is diffuse, leading to limited performance under non-diffuse field conditions. 

Aiming to overcome the limitations of signal-independent binaural encoding, recent studies incorporated signal-dependent or data-driven approaches. These are typically based on sound field parametrization into distinct sources and ambient components \cite{Pulkki:DirectionalAudioCoding:2006,Politis:COMPASSCodingMultidirectional:2018,Hold:OptimizingHigherorderDirectional:2023,McCormack:SPARTACOMPASSRealtime:2019}, showing binaural reproduction with high perceptual quality due to the focus on reproduction the direct sound from sources. 
%
These methods were recently incorporated into the BSM framework, where the ambient component is reproduced using the signal-independent BSM \cite{Hermon:BinauralSignalMatching:2022, Berger:PerformanceAnalysisBinaural:2022, Berger:PerformanceRobustnessSignaldependent:2026}. 
Notably, Berger~\cite{Berger:PerformanceRobustnessSignaldependent:2026} demonstrated that this signal-dependent BSM approach offers a robust yet perceptually superior alternative to the signal-independent BSM, showing advantages under listener head rotations and high direct-to-reverberant ratio (DRR) conditions. 
Despite the improved performance of this signal-dependent BSM framework and the other signal-dependent approaches referenced above, they strongly depend on reliable estimation of the direction-of-arrival (DOA) of sound sources and their source signal, which may limit performance under complex or challenging acoustic environments with multiple sources and reverberation.

This paper proposes a novel
framework that utilizes the proven advantages of the signal-dependent BSM approach, while avoiding direct estimation of the acoustic scene parameters. This is achieved by incorporating a Deep Neural Network (DNN) to calculate
the parameters of a directional weighting, denoted as the Field-of-View (FoV), which is then integrated into the BSM filters. 
FoV-based approaches have been previously incorporated in binaural reproduction for near-field BSM
\cite{Goldring:BinauralSignalMatching:2026}, audio zoom \cite{fernandez2024binaural}, and moving speakers \cite{Mittal:MixtureofExpertsFrameworkFieldofView:2025}, demonstrating the power of the FoV weighting approach.  
However, none have studied the use of FoV with BSM in a deep-learning framework. 
By learning complex spatial mappings end-to-end, the proposed approach
aims to equip BSM with a signal-dependent FoV enhancement, therefore avoiding the need for explicit sound field parameter estimation, while retaining the performance benefits of signal-dependent BSM. Evaluation using numerical simulations and a listening test demonstrates performance on a par with signal-dependent methods when estimation accuracy is high, improved robustness compared to these methods when conditions become challenging, and superior performance over the signal-independent BSM.

\section{Mathematical Background}
\label{sec:Mat_Back}
This section presents the mathematical background underlying
the proposed framework, namely the microphone array signal model, the binaural signal model, and both the BSM and FoV formulations.

\subsection{Microphone Array Signal Model}
\label{subsec:micmodel}
Consider an $M$-element microphone array positioned at the origin of a spherical coordinate system. The captured sound field is assumed to comprise $Q$ far-field sources, each generating a plane wave arriving from direction $\{\Omega_q = (\theta_q, \phi_q)\}_{q=1}^Q$. The elevation angle $\theta\in[-\frac{\pi}{2},\frac{\pi}{2}]$ is measured from the horizontal plane, with positive angles representing directions above this plane, while the azimuth angle $\phi\in[-\pi,\pi]$ is measured in the $xy$-plane from the positive $x$ axis. With the wavenumber denoted as
$k$, the sound pressure measured by the array is described by the narrowband model \cite{VanTrees:OptimumArrayProcessing:2002} as:
\begin{equation}
    \mathbf{x}(k)=\mathbf{V}(k)\mathbf{s}(k)+\mathbf{n}(k).
    \label{eqn:model}
\end{equation}

In this expression ${\mathbf{x}(k)=[x_1(k),...,x_M(k)]^T}$ is the measured microphone-signal vector, $\mathbf{V}(k)=[\mathbf{v}(k,\Omega_1),...,\mathbf{v}(k,\Omega_Q)]$ is an $M\times Q$ complex matrix whose columns $\mathbf{v}(k,\Omega_q)=[v_1(k,\Omega_q),...,v_M(k,\Omega_q)]$ denote the array steering vectors corresponding to the $q$-th source and all $M$ microphones, ${\mathbf{s}(k)=[s_1(k),...,s_Q(k)]^T}$ is the source-signal vector, and $\mathbf{n}(k)$ is an additive-noise vector.

\subsection{Binaural Signal Model}
\label{subsec:binsig}
Under the same assumption of $Q$ far-field sources, the binaural pressure signals at the listener's ears, with the listener's head center co-located with the array center, are modeled through the convolution of each source signal with its corresponding HRTF:

%
\begin{equation}
    p^{l,r}(k) = [\mathbf{h}^{l,r}(k)]^T \mathbf{s}(k),
\end{equation}
where $\textbf{h}^{l,r}(k)=[h^{l,r}(k, \Omega_1),\dots, h^{l,r}(k,\Omega_Q)]^T$ is a $Q$-length vector holding the transfer functions from each source direction $\Omega_q$ to the listener's left ($l$) or right ($r$) ear. These narrowband signals can subsequently be transformed into the time domain for headphone-based binaural playback.

\subsection{Binaural Signal Matching (BSM)}
\label{subsec:BSM}
Building on the signal models introduced above, the BSM method estimates the target binaural signal $\hat{p}^{l,r}(k)$ by applying a filter to the microphone signals, analogous to beamforming. The estimated signal is then given by:
\begin{equation}
    \hat{p}^{l,r}(k)=[\mathbf{c}^{l,r}(k)]^H\mathbf{x}(k),
    \label{eqn:bin_sig}
\end{equation}
where $\mathbf{c}^{l,r}(k)$ is an $M\times1$ complex vector of filter coefficients, and $(.)^H$ denotes the Hermitian operator. The coefficients $\mathbf{c}^{l,r}(k)$ are obtained by minimizing the mean-squared error:
\begin{equation}
    \mathbf{c}^{l,r}_{opt}(k)= \arg\min_{\mathbf{c}} 
    \mathbb{E}[|\hat{p}^{l,r}(k)-p^{l,r}(k)|^2]
    \label{eqn:bsm_err2}
\end{equation}
with $\mathbb{E}[\cdot]$ denoting the expectation operator.
Minimizing the error in Eq.~(\ref{eqn:bsm_err2}) yields the following solution~\cite{Madmoni:DesignAnalysisBinaural:2024}:
\begin{equation}
        \mathbf{c}^{l,r}_{opt}=(\mathbf{V}\mathbf{R_s}\mathbf{V}^H+\mathbf{R_n})^{-1}\mathbf{V}\mathbf{R_s}[\mathbf{h}^{l,r}]^*
    \label{eqn:6}
\end{equation}
where $\mathbf{R_s}=E[\mathbf{s}\mathbf{s}^H]$, $\mathbf{R_n}=E[\mathbf{n}\mathbf{n}^H]$ denote the source and noise covariance matrices, respectively. The development of the BSM in \cite{Madmoni:DesignAnalysisBinaural:2024} assumed that the source signals are independent and identically distributed (iid), which is similar to the assumption of a diffuse sound field. Also, the noise is modeled as spatially white, as typically assumed for sensor noise. Under these assumptions, the covariance matrices reduce to $\mathbf{R}_s = \sigma_s^2 \mathbf{I}_Q$, $\mathbf{R}_n = \sigma_n^2 \mathbf{I}_M$, where $\sigma_s^2$ and $\sigma_n^2$ represent the source and noise variances, respectively, and $\mathbf{I}_Q$, $\mathbf{I}_M$ are the corresponding $Q \times Q$ and $M \times M$ identity matrices. Substituting these into Eq. (\ref{eqn:6}) further simplifies the BSM filter to:
\begin{equation}
    \mathbf{c}^{l,r}_{BSM}=(\mathbf{V}\mathbf{V}^H+\frac{1}{SNR}\mathbf{I}_M)^{-1}\mathbf{V}[\mathbf{h}^{l,r}]^*
    \label{eqn:opt_BSM}
\end{equation}
where $SNR=\frac{\sigma_s^2}{\sigma_n^2}$.

To mitigate the degradation of spatial cues at high frequencies, where BSM accuracy decreases and the binaural error grows, a Magnitude Least Square (MagLS) modification of BSM was introduced in
\cite{Setsompop:MagnitudeLeastSquares:2008,Deppisch:EndtoEndMagnitudeLeast:2021a,Madmoni:DesignAnalysisBinaural:2024}, in which only the magnitude of the error in Eq. (\ref{eqn:bsm_err2}) is minimized. 
This choice is motivated by the perceptual observation that, at high frequencies, the Interaural Level Difference (ILD) carries greater perceptual salience than the Interaural Time Difference (ITD) \cite{Brughera:HumanInterauralTime:2013a, Macpherson:ListenerWeightingCues:2002a}. 
 
\subsection{Field of View (FoV) BSM}
\label{subsec:FoV}
In applications of BSM to wearable arrays, sound sources such as human speakers may be positioned in front of the person wearing the array device, therefore limited in space to a specified FoV. This assumption has been previously incorporated in a FoV-BSM method to reduce binaural error by increasing the importance of FoV directions \cite{Goldring:BinauralSignalMatching:2026}. Within such a framework, the FoV is defined as the angular region in spherical coordinates $(\theta,\phi)$ around a front-looking axis, and a corresponding spatial weighting function $w(\theta, \phi)$ is introduced to weight the steering matrix and HRTF vector:

\begin{equation}
   w(\theta, \phi) = 
   \begin{cases} 
      1, & \text{if } (\theta, \phi) \in \text{FoV}, \\ 
      \beta, & \text{otherwise}. 
   \end{cases}
   \label{eqn:wFoV}
\end{equation}
Here $\beta$ is a small positive constant that balances binaural-error minimization inside and outside the FoV. The weighted steering matrix and HRTF vector can now be defined as:
\begin{equation}
    \tilde{\mathbf{V}} = \mathbf{V}\mathbf{W}
    \label{eqn:fov1}
\end{equation}
\begin{equation}
    \tilde{\mathbf{h}}^{l,r} = \mathbf{W}\mathbf{h}^{l,r},
    \label{eqn:fov2}
 \end{equation}
with $\mathbf{W}$ = diag($w(\theta_{1}, \phi_{1}), ..., w(\theta_{Q},\phi_{Q}))$ holding the weights for the $Q$ directions. Substituting $\mathbf{\tilde{V}}$ and $\mathbf{\tilde{h}}^{l,r}$ in Eq. (\ref{eqn:opt_BSM}) yields the FoV-BSM filter, which forms the basis for the signal-dependent formulation developed in the following sections.

\section{Proposed Method}
\label{sec:Prop_Meth}
This section presents a novel deep learning approach for inferring FoV parameters from the microphone signals, yielding a signal-dependent FoV-BSM. 

\begin{figure*}[t]
\centering
\includegraphics[width=\textwidth]{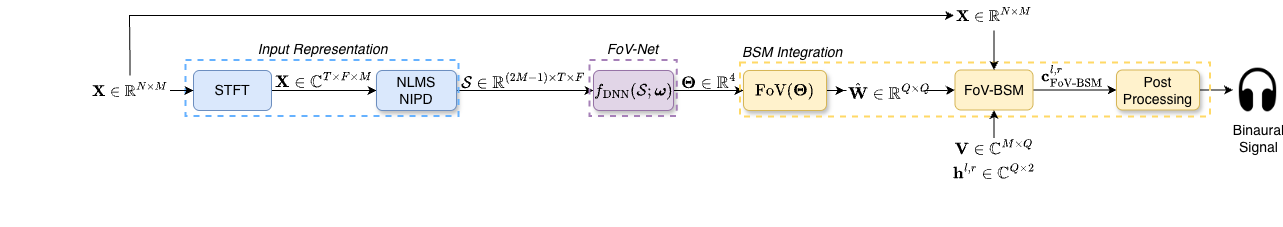}
\caption{Overview of the proposed neural FoV-BSM-Net pipeline. The pipeline consists of three main stages: (i) Input representation, which extracts spatial cues from the multichannel signal; (ii) A DNN, which predicts FoV parameters to generate the directional weighting matrix $\mathbf{\hat{W}}$; and (iii) BSM integration, which computes the final binaural filters by combining the data-driven FoV weights with the model-based BSM back-end.}
\label{fig:pipeline}
\end{figure*}

\subsection{The proposed approach}
\label{subsec:prop_aprch}
The proposed approach is motivated by a recent study that demonstrated the benefit of the FoV-BSM for both near-field and far-field sources \cite{Goldring:BinauralSignalMatching:2026}. However, the study used a fixed FoV assuming the target speaker is positioned in front of a listener equipped with a wearable microphone array. The proposed approach extends the FoV to be signal-dependent, therefore tailoring the FoV to the position of the target speaker.

One possible approach is to estimate the target speaker direction using source localization methods such as GCC-PHAT~\cite{Knapp:GeneralizedCorrelationMethod:1976}, SRP~\cite{Dibiase:HighaccuracyLowlatencyTechnique:2000}, and real-time multi-source trackers~\cite{Pavlidi:RealTimeMultipleSound:2013a}. However, this approach has limitations. First, errors in DOA estimation, due, for example, to reverberation and noise, may reduce performance; this limitation can be alleviated to some extent by recent deep-learning-based DOA estimation and source-counting methods~\cite{Adavanne:DirectionArrivalEstimation:2018,Chakrabarty:MultiSpeakerDOAEstimation:2019b,Nguyen:RobustSourceCounting:2020b}. Second, FoV-BSM also requires the estimation of the FoV directional region (width and height) which is not provided by classical DOA estimation methods.  

The proposed approach is therefore more direct: FoV estimation is formulated as a signal-dependent regression problem in which a DNN maps a short-time microphone-array segment to four parameters that fully describe the FoV. The FoV is parameterized by a centre $(\theta_c,\phi_c)$ in spherical coordinates and an angular aperture $(\Delta_\theta,\Delta_\phi)$, leading to the parameter vector
\begin{equation}
    \boldsymbol{\Theta} \equiv [\theta_c,\,\phi_c,\,\Delta_\theta,\,\Delta_\phi]^{T},
\end{equation}
which defines the FoV region
\begin{equation}
    \mathrm{FoV}(\boldsymbol{\Theta}) = \big\{(\theta,\phi)\,:\,|\theta-\theta_c|\le\Delta_\theta,\,|\phi-\phi_c|\le\Delta_\phi\big\},
    \label{eqn:fovregion}
\end{equation}
and, through Eq.~(\ref{eqn:wFoV}), the weighting function $w(\theta,\phi)$.

The full pipeline, illustrated in Fig.~\ref{fig:pipeline} and denoted as FoV-BSM-Net, runs in three stages. First, the microphone signals are short-time Fourier transformed (STFT) into $X_m(t,f)$, $m=1,\dots,M$, and combined into a SALSA-Lite-style spatial representation $\mathcal{S}$~\cite{ThoNguyen:SALSALiteFastEffective:2022}. Second, a CRNN maps $\mathcal{S}$ to the FoV parameter vector $\boldsymbol{\Theta}$. Third, $\boldsymbol{\Theta}$ defines a diagonal weighting matrix $\hat{\mathbf{W}}$ that is integrated into the BSM filter design. The three stages are detailed in the following subsections.

\subsection{Input representation}
\label{subsec:dnn_input}
Let $X_m(t,f)\in\mathbb{C}$ denote the STFT of the $m$-th microphone signal at time frame $t=0,\dots,T-1$ and frequency bin $f=0,\dots,F-1$, for $m=1,\dots,M$, with microphone $1$ arbitrarily chosen as the reference. Following SALSA-Lite~\cite{ThoNguyen:SALSALiteFastEffective:2022}, two complementary feature maps are stacked along the channel axis: a multi-channel log-magnitude spectrogram and a normalized inter-channel phase-difference (NIPD) feature.

The first feature, the per-channel normalized log-magnitude spectrogram (NLMS), is defined as
\begin{equation}
    L_m(t,f) = \frac{\log\!\big(|X_m(t,f)|\big)-\mu_m}{\sigma_m},
    \label{eqn:logspec}
\end{equation}
with $(\mu_m,\sigma_m)$ the channel-wise mean and standard deviation computed over the $T\!\times\!F$ grid.
The second is the NIPD with respect to the reference microphone,
\begin{equation}
    \Lambda_m(t,f)
    = \frac{1}{\pi}\,\arg\!\big(X_m(t,f)\,X_1^*(t,f)\big),
    \label{eqn:nipd}
\end{equation}
where $\arg(\cdot)$ returns the phase difference in the range $[-\pi,\pi]$, such that $\Lambda_m(t,f)\in[-1,1]$ by construction. This dimensionless form differs from the original SALSA-Lite NIPD~\cite{ThoNguyen:SALSALiteFastEffective:2022}, which scales the wrapped phase by $c/(2\pi f)$. Stacking the $M$ log-magnitudes and the $M-1$ NIPDs along the channel axis yields the input tensor $\mathcal{S}\in\mathbb{R}^{(2M-1)\times T\times F}$ presented to the network.

\subsection{Network architecture}
\label{subsec:dnn_model}
\begin{figure}[t]
\centering
\includegraphics[width=0.95\columnwidth,height=15cm]{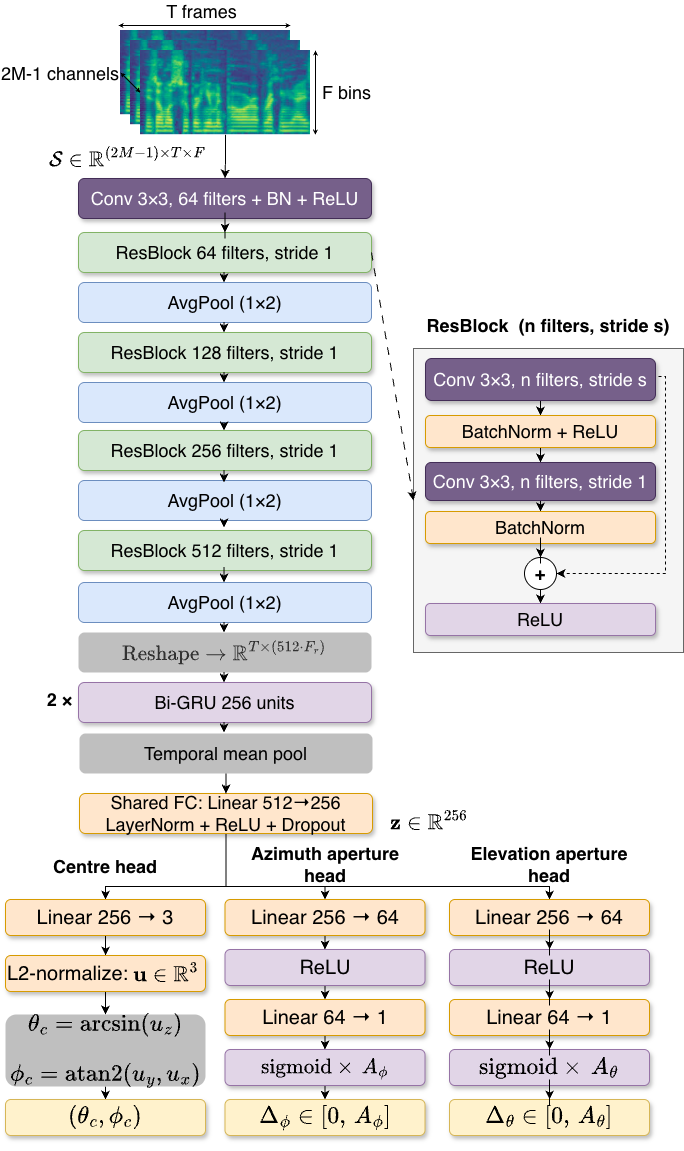}
\caption{Internal structure of the FoV-Net front-end. The architecture follows a CRNN topology that maps the input tensor $\mathcal{S}$ to the FoV parameters $\boldsymbol{\Theta}$. A ResNet-style backbone of stacked residual blocks serves as the convolutional encoder, followed by a two-layer Bi-GRU for temporal modeling. After a shared fully-connected layer, the model branches into three task-specific regression heads that jointly estimate the FoV centre and its angular apertures.}
\label{fig:FoV_Model}
\end{figure}

A Convolutional Recurrent Neural Network (CRNN) is adopted for its effectiveness in capturing local spectro-spatial structure via convolutional layers and modeling temporal dependencies through recurrent layers~\cite{Adavanne:SoundEventLocalization:2019,Adavanne:DirectionArrivalEstimation:2018,ThoNguyen:SALSALiteFastEffective:2022,Chakrabarty:MultiSpeakerDOAEstimation:2019b}. Fig.~\ref{fig:FoV_Model} shows the network architecture used in this work. Following the design of the SELD network in SALSA-Lite~\cite{ThoNguyen:SALSALiteFastEffective:2022}, the model consists of a ResNet-style convolutional backbone, a two-layer bidirectional GRU, and a shared fully-connected (FC) projection that feeds three regression heads. The number of input channels in the first convolutional layer is set to $2M-1$, matching the channel dimension of $\mathcal{S}$ defined in Sec.~\ref{subsec:dnn_input}.

The convolutional backbone extracts hierarchical spectro-spatial features from $\mathcal{S}$ while progressively reducing the frequency dimension and preserving the time resolution. Its output is processed by the bidirectional GRU, which models the temporal evolution of these features. A temporal mean pooling produces a single utterance-level embedding $\mathbf{z}$, which the three heads map jointly to the FoV parameters $\boldsymbol{\Theta}$.

Following the Cartesian-regression DOA branch of~\cite{ThoNguyen:SALSALiteFastEffective:2022}, the \emph{centre head} outputs a unit-norm vector $\mathbf{u}\in\mathbb{R}^{3}$, from which the centre angles are obtained by converting $\mathbf{u}$ from Cartesian to spherical coordinates. The two \emph{aperture heads} use a sigmoid scale factor ($A_\phi{=}180^{\circ}$, $A_\theta{=}90^{\circ}$), so that $\Delta_\phi$ and $\Delta_\theta$ are bounded by construction. As a result, every prediction $\boldsymbol{\Theta}$ corresponds to a valid FoV region, and hence to a well-posed FoV-BSM filter in Eq.~(\ref{eqn:FoV-BSM}). All remaining architectural details are reported in Fig.~\ref{fig:FoV_Model} and in the open-source implementation accompanying this paper.

\subsection{Integration with BSM framework}
\label{subsec:integBSM}
Given the predicted FoV parameters $\boldsymbol{\Theta}$, the weighting function $w(\theta,\phi)$ of Eq.~(\ref{eqn:wFoV}) is evaluated on the $Q$ steering directions used by the BSM design. To enable end-to-end training, the hard indicator in Eq.~(\ref{eqn:wFoV}) is replaced by a differentiable soft mask $\hat{w}(\theta,\phi;\boldsymbol{\Theta})\in[\beta,1]$. The weighting function defines the diagonal matrix $\hat{\mathbf{W}}$, which is substituted into Eqs.~(\ref{eqn:fov1})--(\ref{eqn:fov2}) and then into Eq.~(\ref{eqn:opt_BSM}), yielding the signal-dependent FoV-BSM filter
\begin{equation}
    \mathbf{c}^{l,r}_{\text{FoV-BSM}}
    = \left(\mathbf{V}\hat{\mathbf{W}}^{2}\mathbf{V}^{H}+\lambda\mathbf{I}_{M}\right)^{-1}
      \mathbf{V}\hat{\mathbf{W}}^{2}\,[\mathbf{h}^{l,r}]^{*},
    \label{eqn:FoV-BSM}
\end{equation}
where the Tikhonov regularization constant $\lambda$ plays the role of the $1/\mathrm{SNR}$ term in Eq.~(\ref{eqn:opt_BSM}). Since every step from $\boldsymbol{\Theta}$ to $\mathbf{c}^{l,r}_{\text{FoV-BSM}}$ is differentiable, the network can be trained end-to-end with a binaural-domain objective, described in the next section.

\section{Error Measures}
\label{sec:Err_Meas}
This section presents the metrics used in the study, first as evaluation measures for comparing binaural reproduction methods (Sec.~\ref{subsec:eval}), and then, with the required adjustments, as training losses for the proposed model (Sec.~\ref{subsec:train_loss}). For clarity, $\hat{p}^{l,r}(t)$ and $p^{l,r}(t)$ denote the estimated and reference binaural signals.

\subsection{Evaluation metrics}
\label{subsec:eval}
Performance is assessed with five complementary metrics: the binaural Normalized Mean-Squared Error (NMSE), the ITD error, the ILD error, the Interaural Channel Level Difference (ICLD) error, and the Interaural Phase Difference (IPD) error.

\textit{1) Binaural NMSE:}
The NMSE is computed in the STFT domain, in two frequency-dependent regions, motivated by the duplex theory of binaural localization~\cite{McAlpine:CreatingSenseAuditory:2005} and consistent with the MagLS criterion used in the BSM filter design (Sec.~\ref{subsec:BSM}). The full complex error is preserved at low frequencies, while only the magnitude error is evaluated above $f_c=1.5\,$kHz, where phase cues lose perceptual salience,
\begin{equation}
    \epsilon_{\text{NMSE}}^{l,r}(f) =
    \begin{cases}
        \dfrac{\sum_{t}|\hat{p}^{l,r}(t,f)-p^{l,r}(t,f)|^{2}}{\sum_{t}|p^{l,r}(t,f)|^{2}}, & f<f_c,\\[6pt]
        \dfrac{\sum_{t}\big||\hat{p}^{l,r}(t,f)|-|p^{l,r}(t,f)|\big|^{2}}{\sum_{t}|p^{l,r}(t,f)|^{2}}, & f\geq f_c,
    \end{cases}
    \label{eqn:nmse}
\end{equation}
where $f$ indexes the STFT frequency bins, and the summation over time is performed over a selected range of frames. 

\textit{2) ITD:}
The ITD is computed via the inter-aural cross-correlation (IACC) following the perceptually-motivated definition of~\cite{Andreopoulou:IdentificationPerceptuallyRelevant:2017}. The binaural signals are first low-pass filtered at $1.5$kHz, in line with the frequency range relevant for binaural time-difference perception, and the ITD is then obtained as
\begin{equation}
    ITD(\Omega) = \arg\max_{\tau}\Big|\sum_{t=0}^{T-\tau-1}p^{r}(t)\,p^{l}(t+\tau)\Big|,
    \label{eqn:ITD}
\end{equation}
having the DOA denoted as $\Omega$, and $T$ as the number of samples. $\widehat{ITD}(\Omega)$ is similar but calculated for the estimated binaural signals. The corresponding error is its absolute deviation from the reference,
\begin{equation}
    \epsilon_{\text{ITD}}(\Omega) = \big|ITD(\Omega) - \widehat{ITD}(\Omega)\big|.
    \label{eqn:ITDerr}
\end{equation}

\textit{3) ILD:}
The ILD is computed over equivalent rectangular bandwidth (ERB) filter bands following~\cite{Xie:HeadrelatedTransferFunction:2013},
\begin{equation}
    ILD(f_c,\Omega) = 10\log_{10}\frac{\sum_{f=0}^{f_c^{\max}}|G(f,f_c)|^{2}|p^{l}(f)|^{2}}{\sum_{f=0}^{f_c^{\max}}|G(f,f_c)|^{2}|p^{r}(f)|^{2}},
    \label{eqn:ILD}
\end{equation}
where $G(f,f_c)$ is the ERB filter centered at $f_c$ and $f_c^{\max}$ denotes its upper frequency limit. $\widehat{ILD}(f_c,\Omega)$ is defined similarly but for the estimated binaural signals. The ILD error is then the mean absolute deviation across the $N_b$ bands of interest,
\begin{equation}
    \epsilon_{\text{ILD}}(\Omega) = \frac{1}{N_b}\sum_{f_c}\big|ILD(f_c,\Omega) - \widehat{ILD}(f_c,\Omega)\big|.
    \label{eqn:ILDerr}
\end{equation}

\textit{4) ICLD error:}
The Interaural Channel Level Difference at direction $\Omega$ and frequency $f$ is defined as
\begin{equation}
    \mathrm{ICLD}(f,\Omega) = 10\log_{10}\frac{|h^{l}(f,\Omega)|^{2}}{|h^{r}(f,\Omega)|^{2}},
    \label{eqn:icld}
\end{equation}
with $\widehat{\mathrm{ICLD}}(f,\Omega)$ defined identically from $\hat{h}^{l,r}(f,\Omega)$. The ICLD error is the absolute deviation between estimated and reference:
\begin{equation}
    \epsilon_{\mathrm{ICLD}}(f,\Omega) = \big|\mathrm{ICLD}(f,\Omega)-\widehat{\mathrm{ICLD}}(f,\Omega)\big|.
    \label{eqn:icld_err}
\end{equation}

\textit{5) IPD error:}
The Interaural Phase Difference at direction $\Omega$ and frequency $f$ is defined as
\begin{equation}
    \mathrm{IPD}(f,\Omega) = \arg\!\big(h^{l}(f,\Omega)\,[h^{r}(f,\Omega)]^{*}\big),
    \label{eqn:ipd}
\end{equation}
with $\widehat{\mathrm{IPD}}(f,\Omega)$ defined analogously from $\hat{h}^{l,r}(f,\Omega)$. The IPD error is the absolute deviation between estimated and reference phases, wrapped to the principal interval $(-\pi,\pi]$ and reported in degrees:
\begin{equation}
    \epsilon_{\mathrm{IPD}}(f,\Omega) = \big|\mathrm{IPD}(f,\Omega)-\widehat{\mathrm{IPD}}(f,\Omega)\big|.
    \label{eqn:ipd_err}
\end{equation}
The per-direction errors $\bar{\epsilon}_{\mathrm{ICLD}}(\Omega)$ and $\bar{\epsilon}_{\mathrm{IPD}}(\Omega)$ are obtained by averaging $\epsilon_{\mathrm{ICLD}}(f,\Omega)$ and $\epsilon_{\mathrm{IPD}}(f,\Omega)$ over frequency, on the same bands used by the ILD and ITD errors: $[f_c,8]\,$kHz for ICLD, and $[0,f_c]\,$kHz for IPD.

\subsection{Training loss}
\label{subsec:train_loss}
The network is supervised with a three-term composite loss, evaluated per mini-batch,
\begin{equation}
    \mathcal{L} = \lambda_{\text{DOA}}\mathcal{L}_{\text{DOA}} + \lambda_{\text{ILD/ITD}}\mathcal{L}_{\text{ILD/ITD}} + \lambda_{\text{NMSE}}\mathcal{L}_{\text{NMSE}},
    \label{eqn:loss_total}
\end{equation}
targeting spatial accuracy, perceptual fidelity, and spectral distortion, respectively.
Each term is built directly from the evaluation metrics of Sec.~\ref{subsec:eval}, with the small adjustments required for end-to-end training detailed below.

\textit{1) DOA loss ($\mathcal{L}_{\text{DOA}}$):}
The centre angles predicted by the FoV head are supervised against the ground-truth direction of the dominant source,
\begin{equation*}
    \mathcal{L}_{\text{DOA}} = (\theta_c - \hat{\theta}_c)^{2} + (\phi_c - \hat{\phi}_c)^{2}.
\end{equation*}
This anchors the FoV centre during training.

\textit{2) ILD/ITD loss ($\mathcal{L}_{\text{ILD/ITD}}$):}
The ILD and ITD measures of Sec.~\ref{subsec:eval} are adapted in two ways for training. First, the discrete $\arg\max$ in Eq.~(\ref{eqn:ITD}) is replaced by a differentiable soft-argmax,
\begin{equation*}
    \tau_{\text{ITD}} = \sum_{\tau=-\tau_{\max}}^{\tau_{\max}}\tau\,\mathrm{softmax}\!\Big(\Big|\sum_{t}p^{r}(t)\,p^{l}(t+\tau)\Big|\Big),
\end{equation*}
with $\tau_{\max}=50$ samples ($3.1$ms), leading to $\mathcal{L}_{\text{ITD}}=|\tau_{ITD}-\hat{\tau}_\text{ITD}|$ with $\hat{\tau}_{ITD}$ denoting the ITD computed from the estimated binaural signals, and $\tau_{ITD}$ from the reference binaural signals. The latter is evaluated on the \emph{binaural transfer function} of the system at the ground-truth source direction $\Omega_c=(\theta_c,\phi_c)$, as detailed below. First, the FoV-BSM filter of Eq.~(\ref{eqn:FoV-BSM}) is combined with the array steering vector at $\Omega_c$,
\begin{equation}
    \hat{h}^{l,r}(\Omega_c) = \big[\mathbf{c}^{l,r}_{\text{FoV-BSM}}(k)\big]^{H}\,\mathbf{v}(k,\Omega_c),
    \label{eqn:hrtf_eff}
\end{equation}
yielding the estimated HRTF, which is compared against the reference HRTF $h^{l,r}(\Omega_c)$ from the dataset of Sec.~\ref{subsec:binsig}. Then, both pairs are converted to the Head-Related Impulse Responses (HRIR) by the inverse Fast Fourier Transform (FFT) and substituted for $(\hat{p}^{l,r},p^{l,r})$ in Eqs.~(\ref{eqn:ITD})--(\ref{eqn:ILDerr}).

The ILD is computed from Eq. (\ref{eqn:ILD}) by averaging over $K{=}14$ ERB-spaced Gammatone bands in the range $[1.5{-}8]\,$kHz. The combined perceptual loss was finally computed as:
\begin{equation*}
    \mathcal{L}_{\text{ILD/ITD}} = w_{\text{ILD}}\,\mathcal{L}_{\text{ILD}} + w_{\text{ITD}}\,\mathcal{L}_{\text{ITD}}.
\end{equation*}
Computing the loss from the HRTF rather than the binaural signal decouples the loss from the source signal content and the room response, aiming to lead to a cleaner training process.

\textit{3) Binaural NMSE ($\mathcal{L}_{\text{NMSE}}$):}
The NMSE training loss is computed directly from Eq.~(\ref{eqn:nmse}), averaged across frequencies and ears and reported in dB. \newline

Although not used in training, IPD/ICLD are essential for per-direction evaluation, revealing direction-dependent errors averaged out by broadband ITD/ILD.

\section{Simulation Study}
\label{sec:sim}
This section describes the simulation setup, the dataset, and the methods compared in this study, including FoV-BSM-Net and its baselines.

\subsection{Setup}
\label{subsec:set-up}
The acoustic scenes are generated using a room simulation framework based on the image-source method (ISM)~\cite{Allen:ImageMethodEfficiently:1979}, implemented via the Pyroomacoustics library~\cite{Scheibler:PyroomacousticsPythonPackage:2018}.

Within each simulation, a single speech source and a wearable microphone array are placed. The source is modeled as a point emitter radiating anechoic speech signals, and the receiver is a free-field microphone array configured according to a reduced version of the Project Aria geometry~\cite{Engel:ProjectAriaNew:2023}, denoted as the \emph{Aria-reduced} layout. This configuration consists of five microphones corresponding to forward-facing sensors: two lower-lens, two front-temple, and one nose-bridge location. The array geometry is used consistently both in the room simulation and in the analytical modeling of steering vectors.

The array steering vectors are formulated in the spherical harmonic (SH) domain following Sec.~4.2 in~\cite{Rafaely:FundamentalsSphericalArray:2015}, under far-field and plane-wave assumptions, and are defined over $Q$ directions sampled nearly uniformly on the sphere via a Fibonacci lattice~\cite{Gonzalez:MeasurementAreasSphere:2009}.

The binaural reference signal is generated using measured HRTFs from the Cologne database for the Neumann KU100 manikin~\cite{Bernschutz:SphericalFarField:2013}. The HRTF set is co-located with the array center, with its frontal axis aligned along the positive \(x\)-direction. For each simulated condition, the binaural signal is obtained by applying the HRTFs corresponding to the source DOA, using the same propagation paths as in the array simulation.

\subsection{Dataset}
\label{subsec:dataset}  
Based on the above setup, a dataset of single-source reverberant scenarios is generated using a Monte Carlo procedure. Each scenario includes a fixed $3\,\mathrm{s}$ segment of multichannel audio, sampled at $16\,\mathrm{kHz}$, together with the corresponding binaural reference signal. The acoustic environments are sampled from a set of predefined room configurations representing typical indoor spaces, with dimensions $(5 \times 4 \times 3)$, $(10 \times 8 \times 3.5)$, and $(20 \times 15 \times 6)$ meters. These configurations correspond to reverberation times ($T_{60}$) of $0.30$, $0.61$, and $1.40$ seconds, respectively. For each realization, the room configuration is selected randomly.

The source signal is drawn from the \emph{train-clean-100} subset of LibriSpeech~\cite{Panayotov:LibrispeechASRCorpus:2015} and center-cropped to this fixed duration. The source DOA is sampled from the FoV training grid \(\phi \times \theta\), yielding \(525\) candidate directions per room (see Table~\ref{table:room_param}), while source positions are randomized within the room volume to cover far-field conditions. For each source–receiver configuration, room impulse responses (RIRs) are generated using the ISM. The corresponding acoustic parameters, including \(T_{60}\) and $DRR$, are then derived from the simulated RIRs.

In total, $12{,}000$ independent scenarios (approximately $10\mathrm{h}$ of audio) are generated by randomly sampling room configurations, source positions, DOAs, and speech utterances. The dataset is partitioned into training, validation, and test sets containing $9{,}600$, $1{,}200$, and $1{,}200$ samples, respectively, ensuring balanced coverage of spatial and acoustic conditions across all splits.

\begin{table}[t]
\centering
\caption{Monte~Carlo database parameters.}
\small
\begin{tabular}{ll}
\hline
\textbf{Parameter} & \textbf{Value} \\
\hline
Room dimensions (m)        & $(5{\times}4{\times}3)$, $(10{\times}8{\times}3.5)$, \\
                           & $(20{\times}15{\times}6)$ \\
$T_{60}$ (s)               & $\{0.30,\;0.61,\;1.40\}$ \\
Array position (m)         & $(1.66,\,2,\,1.8)$, $(3.33,\,4,\,1.8)$, \\
                           & $(7.92,\,9.66,\,1.8)$ \\
Source azimuth ($^\circ$)  & $-60\!:\!5\!:\!60$ \\
Source elevation ($^\circ$)& $-30\!:\!3\!:\!30$ \\
Source distance (m)        & $[0.15,\,12.3]$ \\
\hline
\end{tabular}
\label{table:room_param}
\end{table}

\subsection{Methodology}
\label{subsec:sim_meth}
Three methods are evaluated on the same dataset:

\begin{itemize}
    \item \textbf{BSM}: The signal-independent baseline BSM filter of Eq.~(\ref{eqn:opt_BSM}) with uniform spatial weighting.
    \item \textbf{FoV-BSM}: The FoV-weighted BSM of Eqs.~(\ref{eqn:fov1})-(\ref{eqn:fov2}) with a front-facing FoV of fixed angular aperture $(60^{\circ}{\times}20^{\circ})$, selected in Sec.~\ref{subsec:res_sweep}, serving as a fixed FoV baseline.
    \item \textbf{FoV-BSM-Net}: The proposed signal-dependent FoV-BSM of Eq.~(\ref{eqn:FoV-BSM}), with FoV parameters $\boldsymbol{\Theta}$ predicted by the network described in Sec.~\ref{sec:Prop_Meth}.
\end{itemize}

All three methods use the same array geometry, HRTFs, MagLS regime, Tikhonov regularization, and STFT parameters, ensuring that performance differences are attributable to the spatial-focus design rather than to inconsistent processing choices.

The microphone-array signals are processed using an STFT with a Bartlett window of $32\,\mathrm{ms}$ ($512$ samples) and a hop size of $16\,\mathrm{ms}$ ($256$ samples). All filters are designed in the frequency domain on a $512$-point FFT grid, over $Q=400$ directions and with Tikhonov regularization $\lambda=10^{-3}$. MagLS~\cite{Deppisch:EndtoEndMagnitudeLeast:2021a} is applied above the crossover frequency $f_c=1.5\,\mathrm{kHz}$, and the FoV-BSM filter of Eq.~(\ref{eqn:FoV-BSM}) additionally uses an out-of-FoV weight $\beta=0.2$. The NMSE metric of Eq.~(\ref{eqn:nmse}) is evaluated across the full STFT band.

The proposed network is trained on the $9{,}600$-scenario training dataset using the Adam optimizer~\cite{Kingma:AdamMethodStochastic:2014}, with a learning rate of $10^{-3}$, weight decay of $10^{-6}$, and mini-batch size of $32$. Training runs for up to $60$ epochs with early stopping on the validation loss. The composite loss of Eq.~(\ref{eqn:loss_total}) is used throughout training, with outer weights $(\lambda_{\text{DOA}},\lambda_{\text{ILD/ITD}},\lambda_{\text{NMSE}})=(0.01,0.6,0.2)$. Within the ILD/ITD term, the ILD and ITD sub-losses are combined using $(w_{\text{ILD}},w_{\text{ITD}})=(1,3)$.

\section{Simulation Results}
\label{sec:sim_res}
This section presents the simulation results in four parts. First, an ad-hoc search in Sec.~\ref{subsec:res_sweep} selects the FoV baseline used in the rest of the paper. Second, the binaural NMSE is compared in Sec.~\ref{subsec:res_nmse} for the methods of Sec.~\ref{subsec:sim_meth}. Third, the FoV-BSM-Net parameters learned by the network are reported in Sec.~\ref{subsec:res_train}. Finally, the IPD and ICLD errors are analyzed in Sec.~\ref{subsec:res_ipd_icld}.

\subsection{FoV baseline parameter study}
\label{subsec:res_sweep}
The parameters of the FoV-BSM baseline used in the following sections are finalized via an ad-hoc search, and include the out-of-FoV weight $\beta$, the azimuth aperture $\Delta_\phi$, and the elevation aperture $\Delta_\theta$, with the FoV center fixed at the front-facing prior $(\theta,\phi){=}(0,0)$~\cite{Goldring:BinauralSignalMatching:2026}. The selected configuration showing the best performance over a diverse parameter search grid was $(\beta,\Delta_\phi,\Delta_\theta){=}(0.2, 60^{\circ}, 20^{\circ})$.
To confirm that this choice is not fragile, Table~\ref{tab:fov_sweep} reports the sensitivity of binaural NMSE to each parameter around the selected operating point, the other two being kept fixed at the chosen values. The chosen point lies in a locally stable region: performance varies smoothly with each parameter, and the operating point sits within a fraction of a dB of the best-performing value on both ears for $\beta$ and $\Delta_\phi$.

\begin{table}[t]
\centering
\caption{Binaural NMSE improvement [dB] relative to BSM around the operating point \((\beta,\Delta_\phi,\Delta_\theta)=(0.2,60^\circ,20^\circ)\), for the left and right ears, averaged over frequency across \(n=50\) scenarios with \(\mathrm{DRR}>5\) dB. Bold entries mark the selected operating point. BSM error is \((-6.13,-5.46)\) dB for the left and right ears, respectively.}
\setlength{\tabcolsep}{6pt}
\renewcommand{\arraystretch}{1.05}
\begin{tabular}{@{}lccccc@{}}
\hline
\multicolumn{6}{@{}l}{\textit{(a) Out-of-FoV weight $\beta$}} \\
\hline
$\beta$ & $0.1$ & $0.2$ & $0.4$ & $0.6$ & $0.8$ \\
Left  & $-1.42$ & $\mathbf{-1.36}$ & $-0.80$ & $-0.52$ & $-0.41$ \\
Right & $-0.29$ & $\mathbf{-0.32}$ & $-0.03$ & $+0.05$ & $+0.12$ \\
\hline
\multicolumn{6}{@{}l}{\textit{(b) Azimuth aperture $\Delta_\phi$ [$^{\circ}$]}} \\
\hline
$\Delta_\phi$ & $20$ & $45$ & $60$ & $90$ & $120$ \\
Left  & $+1.21$ & $-1.32$ & $\mathbf{-1.36}$ & $-1.02$ & $-0.51$ \\
Right & $+0.52$ & $-0.09$ & $\mathbf{-0.32}$ & $-0.27$ & $+0.29$ \\
\hline
\multicolumn{6}{@{}l}{\textit{(c) Elevation aperture $\Delta_\theta$ [$^{\circ}$]}} \\
\hline
$\Delta_\theta$ & $10$ & $20$ & $45$ & $60$ & $90$ \\
Left  & $-1.00$ & $\mathbf{-1.36}$ & $-1.27$ & $-1.03$ & $-0.84$ \\
Right & $-0.25$ & $\mathbf{-0.32}$ & $-0.73$ & $-0.66$ & $-0.28$ \\
\hline
\end{tabular}
\label{tab:fov_sweep}
\end{table}

This configuration is used as the FoV-BSM baseline for benchmarking FoV-BSM-Net in Secs.~\ref{subsec:res_nmse} and~\ref{subsec:res_ipd_icld}.

\subsection{Binaural NMSE evaluation}
\label{subsec:res_nmse}
This subsection evaluates the performance over the entire test dataset for the three methods. Table~\ref{tab:nmse_band} shows average NMSE values, in dB, over several DRR ranges. 
The table shows that FoV-BSM-Net achieves a $1.0$\,dB improvement over BSM, compared to $0.5$\,dB for the FoV-BSM baseline, under all tested DRR conditions. At high DRR, these values increase to $1.7$\,dB and $0.8$\,dB, respectively, as expected, because the direct sound from the source becomes more dominant within the FoV. 

\begin{figure}[t]
\centering
\includegraphics[width=\columnwidth]{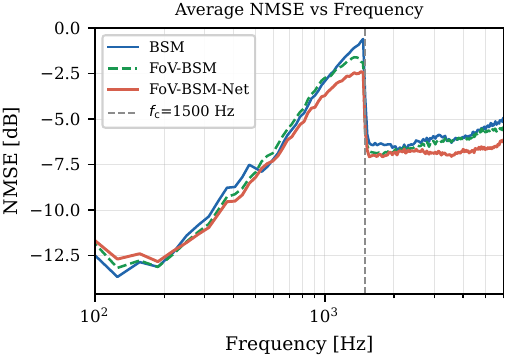}
\caption{Binaural NMSE as a function of frequency, averaged over $N{=}1{,}200$ samples and across both ears. Three filters are compared: BSM (blue), FoV-BSM at $(60^{\circ}{\times}20^{\circ})$ (green dotted), and FoV-BSM-Net (red). The vertical dashed line marks $f_c{=}1.5$\,kHz, separating the complex least-squares and magnitude least-squares design regions.}
\label{fig:nmse_freq_pop}
\end{figure}

Figure~\ref{fig:nmse_freq_pop} presents the frequency-dependent binaural NMSE averaged across the test dataset and over both ears. As shown in the figure, FoV-BSM-Net remains at or below both baselines from approximately $200$\,Hz onward, with an increasing margin over BSM above the crossover frequency $f_c$. On the other hand, the FoV-BSM baseline closely follows BSM throughout the frequency range. 

\begin{table}[t]
\centering
\caption{Test-dataset full-band NMSE [dB], averaged across both ears, with $N$ the number of test samples.}
\small
\setlength{\tabcolsep}{4pt}
\renewcommand{\arraystretch}{1.1}
\begin{tabular}{@{}lcccc@{}}
\hline
\textbf{Method}~\textbackslash~\textbf{DRR [dB]} & $\mathbf{(3,5]}$ & $\mathbf{(5,8]}$ & $\mathbf{{>}8}$ & \textbf{All} \\
\textit{N} & \textit{168} & \textit{159} & \textit{328} & \textit{1200} \\
\hline
BSM                  & $-5.0$ & $-5.2$ & $-6.0$ & $-5.0$ \\
FoV-BSM              & $-5.6$ & $-5.9$ & $-6.8$ & $-5.5$ \\
\textbf{FoV-BSM-Net} & $\mathbf{-6.0}$ & $\mathbf{-6.5}$ & $\mathbf{-7.7}$ & $\mathbf{-6.0}$ \\
\hline
\end{tabular}
\label{tab:nmse_band}
\end{table}

\subsection{FoV-BSM-Net learned parameters}
\label{subsec:res_train}
This subsection reports the FoV parameters learned by FoV-BSM-Net over the test dataset. Table~\ref{tab:doa_span} presents the mean, standard deviation, minimum, and maximum of the centre direction deviations from the source direction, and the learned apertures defined in Sec.~\ref{subsec:prop_aprch}. The table shows that centre direction deviations remain mostly below the training-grid resolution ($5^{\circ}{\times}3^{\circ}$, Table~\ref{table:room_param}), indicating relatively accurate DOA estimation. The learned apertures, on the other hand, are opposite to those selected for FoV-BSM, favoring a wider elevation and a narrower azimuth aperture. This likely reflects the network's confidence in capturing the direct sound, while the wide elevation compensates for residual uncertainty.

\begin{table}[t]
\centering
\caption{FoV parameters - center direction deviation from the source direction, azimuth aperture and elevation aperture, with their mean, standard deviation (Std), and minimum (Min) and maximum values (Max) over the test dataset ($N{=}1{,}200$ samples).}
\small
\setlength{\tabcolsep}{4pt}
\renewcommand{\arraystretch}{1.1}
\begin{tabular}{@{}lcccc@{}}
\hline
\textbf{Parameter ($^\circ$)} & \textbf{Mean} & \textbf{Std} & \textbf{Min} & \textbf{Max} \\
\hline
$|\hat{\theta}_c - \theta_c|$  & $1.3$ & $1.1$ & $0.0$ & $10.0$ \\
$|\hat{\phi}_c - \phi_c|$     & $1.3$ & $1.0$ & $0.0$ & $6.1$  \\
$\hat{\Delta}_\theta$          & $80.4$ & $13.7$ & $32.3$ & $90.0$  \\
$\hat{\Delta}_\phi$            & $35.9$ & $9.3$  & $23.6$ & $70.7$  \\
\hline
\end{tabular}
\label{tab:doa_span}
\end{table}

\subsection{IPD and ICLD error analysis}
\label{subsec:res_ipd_icld}
This subsection evaluates the IPD and ICLD errors of the binaural filters. This evaluation requires a special procedure as detailed next. For this evaluation, a room of dimensions $10{\times}8{\times}3.5\,$m with $T_{60}{=}0.61\,$s and array position of $(3.33, 4, 1.8)\,$m is used, as specified in Table~\ref{table:room_param}. The source direction relative to the array is then varied over the range $\phi\in[-60^{\circ},+60^{\circ}]$ in $1^{\circ}$ steps, and at a fixed elevation $\theta{=}0^{\circ}$ and distance of $0.62\,$m, generating 121 different scenarios. For each scenario, microphone array signals are generated as detailed in Sec.~\ref{subsec:dataset}. Then, the three methods are applied as detailed in Sec.~\ref{subsec:sim_meth}.  
For each method, the estimated binaural transfer function is obtained by applying the filter computed by the method to the array steering vector at the given source direction, as in Eq.~(\ref{eqn:hrtf_eff}). Note that this transfer function represents only the direct sound from the source and is therefore suitable for IPD and ICLD computation; the filters themselves, however, are derived from the reverberant scene.
The band-averaged errors $\bar{\epsilon}_{\mathrm{IPD}}(\Omega)$ and $\bar{\epsilon}_{\mathrm{ICLD}}(\Omega)$ are computed as detailed in Sec.~\ref{subsec:eval} by comparing the estimated transfer function to the anechoic KU100 HRTF at the same direction.

\begin{figure}[t]
\centering
\includegraphics[width=\columnwidth, trim={0 0 0 20pt}, clip]{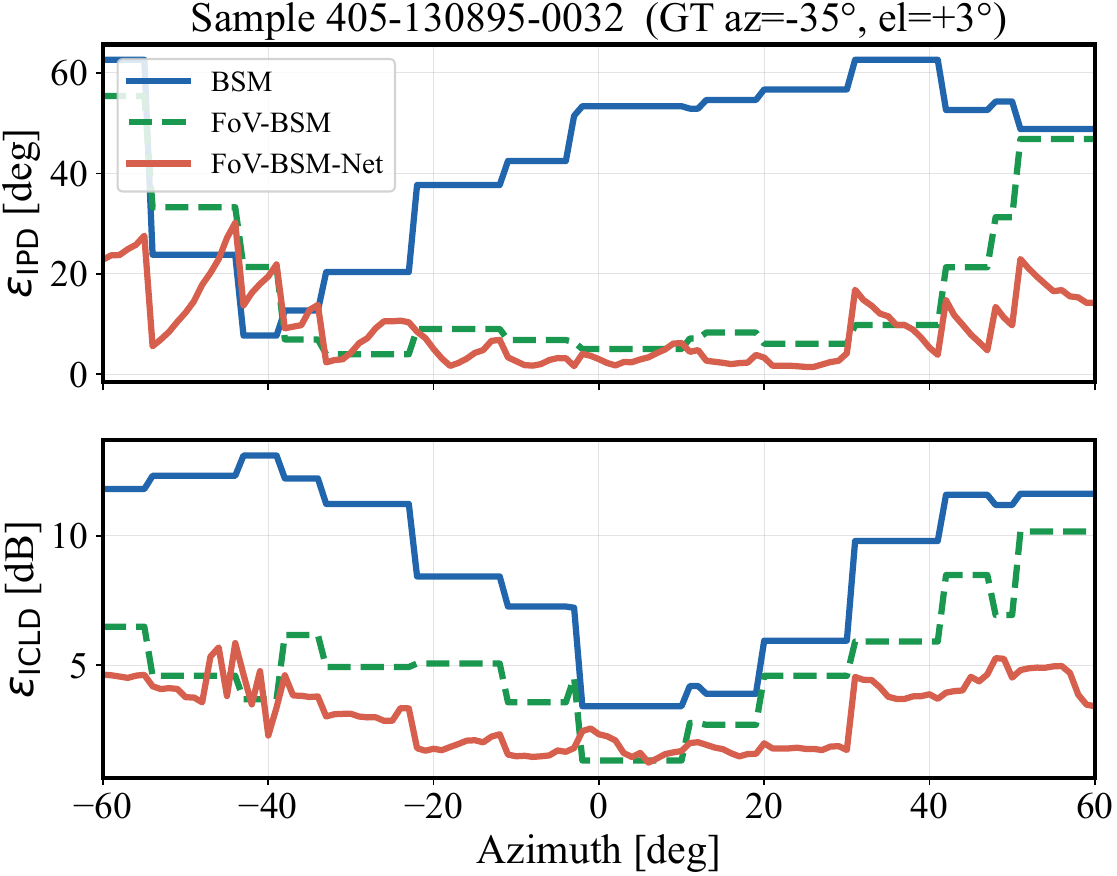}
\caption{IPD (top) and ICLD (bottom) errors for BSM (blue), FoV-BSM-Net (red), and FoV-BSM (green dotted) as a function of source azimuth, as detailed in Sec.~\ref{subsec:res_ipd_icld}. Recall that the FoV-BSM has a FoV with a $60^{\circ}$ aperture in azimuth.}

\label{fig:cues_dynamic}
\end{figure}

Table~\ref{tab:cues} shows the errors averaged across all azimuth directions. The table shows that FoV-BSM-Net achieves the lowest mean errors, reducing ICLD by $5.8$\,dB and IPD by $34.1^{\circ}$ relative to BSM, while FoV-BSM yields smaller reductions of $3.8$\,dB and $26.3^{\circ}$. Fig.~\ref{fig:cues_dynamic} illustrates the ICLD and IPD errors as a function of source direction. The figure shows that the ICLD error for FoV-BSM remains low within $\phi\in[-30^{\circ},+30^{\circ}]$, and increases toward the edges without exceeding BSM. FoV-BSM-Net shows smaller errors, including at wider angles. A similar behavior is also observed for the IPD error, illustrating the overall superior performance of FoV-BSM-Net in maintaining the binaural cues. 

\begin{table}[t]
\centering
\caption{Mean IPD and ICLD errors averaged over the azimuth sweep $\phi\in[-60^{\circ},+60^{\circ}]$ at fixed elevation $\theta{=}0^{\circ}$ and fixed source-to-array distance, for the setup of Sec.~\ref{subsec:res_ipd_icld}.}
\small
\setlength{\tabcolsep}{6pt}
\renewcommand{\arraystretch}{1.1}
\begin{tabular}{@{}lcc@{}}
\hline
\textbf{Method} & $\bar{\epsilon}_{\mathrm{ICLD}}$ [dB] & $\bar{\epsilon}_{\mathrm{IPD}}$ [$^{\circ}$] \\
\hline
BSM                  & $8.9$  & $43.2$ \\
FoV-BSM        & $5.1$  & $16.9$ \\
\textbf{FoV-BSM-Net} & $\mathbf{3.1}$ & $\mathbf{9.1}$ \\
\hline
\end{tabular}
\label{tab:cues}
\end{table}

\section{Listening Experiment}
The simulation results in the previous section offer valuable insights into the objective performance of the different BSM approaches, but they do not fully capture how these differences are perceived by human listeners. This section describes a listening experiment designed to subjectively compare the quality of the proposed method against its baselines.

\subsection{Setup}
\label{sec:exp-setup}
The experimental setup simulates a single point source inside a room, using the same array layout, image-source room simulation, and binaural synthesis as in Sec.~\ref{subsec:set-up}. Speech samples were drawn from the same dataset as in Sec.~\ref{subsec:dataset}, and the FoV-BSM-Net weights were obtained from the training pipeline of Sec.~\ref{subsec:sim_meth}.

The study employed the Multiple Stimuli with Hidden Reference and Anchor (MUSHRA) test~\cite{ITU-RRecommendation:MethodSubjectiveAssessment:2003}, implemented in MATLAB~\cite{MathWorksInc.:MATLABVersion2420:2024b}.
Two scenarios drawn from the dataset distribution of Sec.~\ref{subsec:dataset} were evaluated, jointly spanning diverse acoustic conditions:
\begin{itemize}
    \item \textbf{Scenario 1 (high-DRR)}: male speaker in a small room ($5{\times}4{\times}3$\,m, $T_{60}{=}0.30$\,s), source at $\Omega_d{=}(\theta,\phi){=}(-18^{\circ},60^{\circ})$, $\mathrm{DRR}{=}12.1$\,dB.
    \item \textbf{Scenario 2 (low-DRR)}: female speaker in a medium room ($10{\times}8{\times}3.5$\,m, $T_{60}{=}0.61$\,s), source at $\Omega_d{=}(\theta,\phi){=}(24^{\circ},-50^{\circ})$, $\mathrm{DRR}{=}4.4$\,dB.
\end{itemize}
Room dimensions and $T_{60}$ are consistent with Table~\ref{table:room_param}. Each scenario formed one MUSHRA screen, yielding two screens in total.

\subsection{Methodology}

Binaural signals were generated as described in Secs.~\ref{subsec:sim_meth} and \ref{sec:exp-setup}. Each MUSHRA screen included four test signals: a hidden reference rendered as the same binaural reference used for training (Sec.~\ref{subsec:binsig}), and the three methods described in Sec.~\ref{subsec:sim_meth}.
Twelve subjects, all reporting normal hearing, participated in the study. The MUSHRA screens and signals were presented in randomized order. Participants rated the similarity of each test signal to the reference on overall quality, including both spatial and timbre qualities. Scores ranged from $0$ to $100$, with $100$ indicating that the test signal was indistinguishable from the reference. Headphone compensation filters cited in~\cite{Bernschutz:SphericalFarField:2013} were applied. Prior to the listening test, participants underwent a training stage to familiarize themselves with the scoring procedure.

\subsection{Results}
\label{subsec:listening_results}
A two-way repeated-measures ANOVA (RM-ANOVA) with two within-subject factors, Method (Ref, BSM, FoV-BSM, FoV-BSM-Net) and Scenario, was performed on the collected scores. Mauchly's test indicated a violation of sphericity for the four-level Method factor ($\varepsilon = 0.63$); accordingly, the Greenhouse--Geisser correction was applied. The analysis revealed statistically significant main effects of Method, $F(1.89, 20.81) = 26.5$, $p < .001$, $\eta^{2}_{p} = .71$, and Scenario, $F(1, 11) = 68.6$, $p < .001$, $\eta^{2}_{p} = .86$. Furthermore, the interaction between Method and Scenario was significant, with $F(1.38, 15.16) = 14.7$, $p < .001$, $\eta^{2}_{p} = .57$, indicating that the ranking of the three methods against the reference depends on the acoustic Scenario.

To further examine the Method effect within each Scenario, post-hoc paired $t$-tests contrasted each method against the reference, with Bonferroni correction across the three comparisons per Scenario. In addition, FoV-BSM-Net vs FoV-BSM comparison was performed per Scenario, using Bonferroni-correction. The distribution of participant ratings across Methods and Scenarios is presented as box plots in Fig.~\ref{fig:mushra}. The results can be summarized as follows:

\begin{figure}[t]
\centering
\includegraphics[width=\columnwidth]{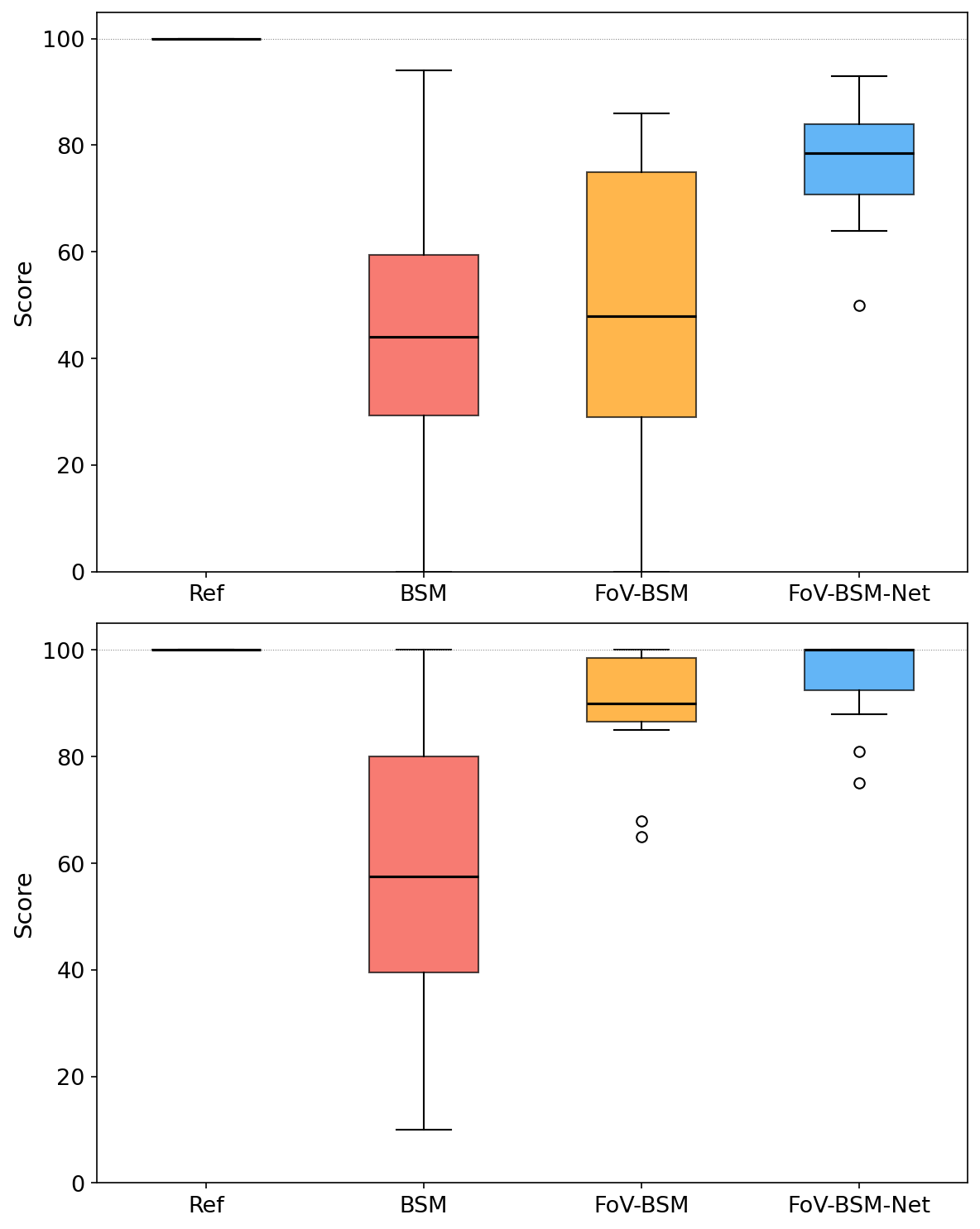}
\caption{Box plots of listening-test scores for the hidden reference (Ref), BSM, FoV-BSM, and FoV-BSM-Net in Scenario~1 (top; small room, high DRR, 12.1\,dB) and Scenario~2 (bottom; medium-size room, low DRR, 4.4\,dB). Boxes span the 25th--75th percentiles, horizontal lines indicate the median, whiskers extend to the most extreme non-outlying scores, and open circles indicate outliers.}
\label{fig:mushra}
\end{figure}

\begin{itemize}
    \item \textbf{Medium room, low DRR (4.4 dB):} the mean differences between the reference and the other methods were $5.3$, $11.3$, and $42.8$ points for FoV-BSM-Net, FoV-BSM, and BSM, respectively, with $p_{corr} = .170$, $.020$, and $< .001$. FoV-BSM-Net was statistically indistinguishable from the reference, while FoV-BSM was rated close to the reference with a small but significant gap. BSM remained clearly worse than the reference. A direct comparison between the two FoV-based methods was not significant, $t(11)=1.33$, $p_{corr}=.420$.

    \item \textbf{Small room, high DRR (12.1 dB):} the mean differences were $23.7$, $52.2$, and $55.2$ points for FoV-BSM-Net, FoV-BSM, and BSM, respectively (all $p_{corr} < .001$). All three methods differed significantly from the reference, but FoV-BSM-Net retained a substantial advantage, while BSM and FoV-BSM degraded to comparable, substantially lower quality. A direct comparison confirmed that FoV-BSM-Net was rated significantly higher than FoV-BSM, $t(11)=3.23$, $p_{corr}=.016$.
\end{itemize}

Overall, the listening-test results corroborate the simulation findings: FoV-BSM-Net matched the reference at low DRR and maintained the highest perceptual quality at high DRR. This confirms the perceptual advantage of adaptive FoV placement, particularly in direct-sound-dominated conditions.

\section{Conclusions}
\label{sec:conclusion}

This paper introduced FoV-BSM-Net, a signal-dependent method for binaural reproduction from wearable microphone arrays that incorporates adaptive FoV estimation into the BSM framework. FoV-BSM-Net consistently improves over BSM, with gains that grow with DRR as direct sound becomes dominant. This improvement is reflected not only in reduced binaural NMSE but also in more accurate interaural cues, indicating better preservation of spatial information. Comparison with a fixed FoV-BSM baseline reveals that fixed FoV weighting accounts for most of the gain when the direct sound aligns with a front-facing aperture, whereas adaptive FoV becomes the primary driver of improvement as the source moves away from the FoV. The learned FoV parameters tend to shift toward a narrower azimuth and wider elevation aperture than the fixed baseline, suggesting that accurate direction estimation reduces the need for broad azimuthal coverage. A listening experiment validated these trends: FoV-BSM-Net was statistically indistinguishable from the reference at low DRR and retained a substantial perceptual advantage over both baselines at high DRR. The evaluation was limited to simulated single-source scenarios; extension to real-world recordings and multi-source conditions are proposed for future work.


\bibliographystyle{IEEEtran}
\bibliography{references}

\end{document}